\documentclass[twocolumn,showpacs,amsmath,amssymb,floatfix,superscriptaddress]{revtex4}
\usepackage{graphicx}% Include figure files 
\usepackage{dcolumn}% Align table columns on decimal point 
\usepackage{bm}% bold math
\usepackage{hyperref} 
\usepackage{latexsym}

\def\b{\beta}

\begin{document}

\title{Ordering of Random Walks: The Leader and the Laggard}
\author{D. ben-Avraham}\email{qd00@clarkson.edu}
\affiliation{Department of Physics, Clarkson University, Potsdam, New York
13699-5820}
\author{B. M. Johnson}
\affiliation{Department of Physics, Clarkson University, Potsdam, New York
13699-5820}
\author{C. A. Monaco}
\affiliation{Department of Physics, Clarkson University, Potsdam, New York
13699-5820}
\author{P.~L.~Krapivsky}\email{paulk@bu.edu}
\author{S. Redner}\email{redner@bu.edu}
\affiliation{Center for BioDynamics, Center for Polymer Studies, and
  Department of Physics, Boston University, Boston, Massachusetts 02215}
\date{\today}

\begin{abstract}
  We investigate two complementary problems related to maintaining the
  relative positions of $N$ random walks on the line: (i) the leader problem,
  that is, the probability ${\cal L}_N(t)$ that the leftmost particle remains
  the leftmost as a function of time and (ii) the laggard problem, the
  probability ${\cal R}_N(t)$ that the rightmost particle never becomes the
  leftmost.  We map these ordering problems onto an equivalent
  $(N-1)$--dimensional electrostatic problem.  From this construction we
  obtain a very accurate estimate for ${\cal L}_N(t)$ for $N=4$, the first
  case that is not exactly soluble: ${\cal L}_4(t)\propto t^{-\beta_4}$, with
  $\beta_4=0.91342(8)$.  The probability of being the laggard also decays
  algebraically, ${\cal R}_N(t)\propto t^{-\gamma_N}$; we derive
  $\gamma_2=1/2, \gamma_3=3/8$, and argue that $\gamma_N\to N^{-1}\ln N$ as
  $N\to\infty$.

\end{abstract}
\maketitle

\section{Introduction} 

Consider $N$ identical and independent random walkers that are initially
located at $x_1(0)<x_2(0)<\cdots< x_{N}(0)$ on an infinite line
\cite{remark}.  There are many interesting questions that can be posed about
the order of the particles.  For example, what is the probability that {\it
  all\/} walkers maintain their relative positions up to time $t$, that is,
$x_1(\tau)<x_2(\tau)<\cdots< x_{N}(\tau)$ for all $0\leq \tau\leq t$?  By
mapping this {\it vicious random walk\/} problem onto the diffusion of a
single effective particle in $N$ dimensions and then exploiting the image
method for the diffusion equation, this ordering probability was found
\cite{Fi84,HF84} to decay asymptotically as $t^{-\alpha_N}$ with
$\alpha_N=N(N-1)/4$.  Many additional aspects of this problem have been
investigated within the rubrics of vicious random walks
\cite{Fi84,FG88,For90,G99,Baik00,KT02,CK02} and ``friendly'' random walks
\cite{GV02,KGV00}.

In this work, we study two related and complementary random walk ordering
problems.  In the ``leader'' problem, we ask for the probability ${\cal
  L}_N(t)$ that the initially leftmost particle in a group of $N$ particles
remains to the left of all the other particles up to time $t$
\cite{N83,Br91,Ke92,Kr96,Re99,Gr02}.  In the ``laggard'' problem, we are
concerned with the probability ${\cal R}_N(t)$ that the initially rightmost
particle from a group of $N$ particles never attains the lead (becomes
leftmost).  These two probabilities ${\cal L}_N(t)$ and ${\cal R}_N(t)$ decay
algebraically
\begin{equation}
\label{LR}
{\cal L}_N(t)\propto t^{-\beta_N}\;,\qquad 
{\cal R}_N(t)\propto t^{-\gamma_N}\;,
\end{equation}
as $t\to\infty$ and our basic goal is to determine the exponents $\beta_N$
and $\gamma_N$.

These ordering problems arise in a variety of contexts.  Physical
applications include wetting phenomena \cite{Fi84,HF84} and three-dimensional
Lorentzian gravity \cite{DL02}.  A more probabilistic application is the
ballot problem \cite{N83}, where one is interested in the probability that
the vote for a single candidate remains ahead of all the other candidates
throughout the counting; this is just a restatement of the leader problem.
Another example is that of the lamb and the lions \cite{Kr96,Re99}, in which
one is interested in the survival of a diffusing lamb in the presence of many
diffusing lions.  In one dimension, a lamb that was initially in the lead
must remain the leader to survive.

For the leader problem, exact results are known for small $N$ only:
$\beta_2=1/2$ and $\beta_3=3/4$ \cite{N83,Av88,FG88,Re01}, while
$\beta_N\to\ln\,(4N)/4$ for large $N$ \cite{Ke92,Kr96,Re99,Gr02}.  This slow
increase arises because adding another particle has little effect on the
survival of the leader when $N$ is large.  For $N\geq 4$, no exact results
are available and one focus of our work is to obtain an accurate estimate of
$\beta_N$ for the case $N=4$.  We accomplish this by mapping the reaction
onto an equivalent electrostatic potential problem due to a point charge
within an appropriately-defined three-dimensional domain.  This mapping
provides both an appealing way to visualize the reaction process and an
accurate estimate of the survival exponent $\beta_4$.  

For the laggard problem, we employ the same method as in the leader problem
to obtain $\gamma_3=3/8$.  We also estimate the asymptotic behavior of ${\cal
  R}_N$ and find $\gamma_N\to N^{-1}\ln N$ as $N\to\infty$.  As is expected,
a laggard in a large population likely remains a laggard.  Therefore the
probability of remaining a laggard decays very slowly with time for large
$N$.

In the next section, we review known results about the leader problem in a
3-particle system.  In Sec.~III, we outline the electrostatic formulation of
the leader problem and then apply it to the 4-particle system in Sec.~IV.  A
numerical solution of the pertinent Laplace equation gives
$\beta_4=0.91342(8)$, a significant improvement over the previously-quoted
estimate $\beta_4\approx 0.91$ \cite{Br91}.  In Sec.~V, we turn to the
laggard problem and give an asymptotic estimate for the exponent $\gamma_N$.
Concluding remarks and some open questions are given in Sec.~VI.

\section{Conventional Approach to the Leader Problem}

We begin by mapping the original problem of $N$ diffusing particles $x_1,
x_2,\ldots, x_{N}$ on the line onto a single effective diffusing particle
located at $\big(x_1,x_2,\ldots,x_{N}\big)$ in $N$-dimensional space.  The
particle order constraints on the line translate to bounding hyperplanes
within which the effective particle is confined \cite{Fi84,Re01}.  The
effective particle is absorbed if it hits one of these boundaries.  For the
leader problem, the explicit shape of these bounding hyperplanes can be
easily worked out for the cases $N=2$ and $N=3$; we will later extend this
analysis to the 4-particle system.

\begin{figure}[ht] 
 \vspace*{0.cm}
 \includegraphics*[width=0.45\textwidth]{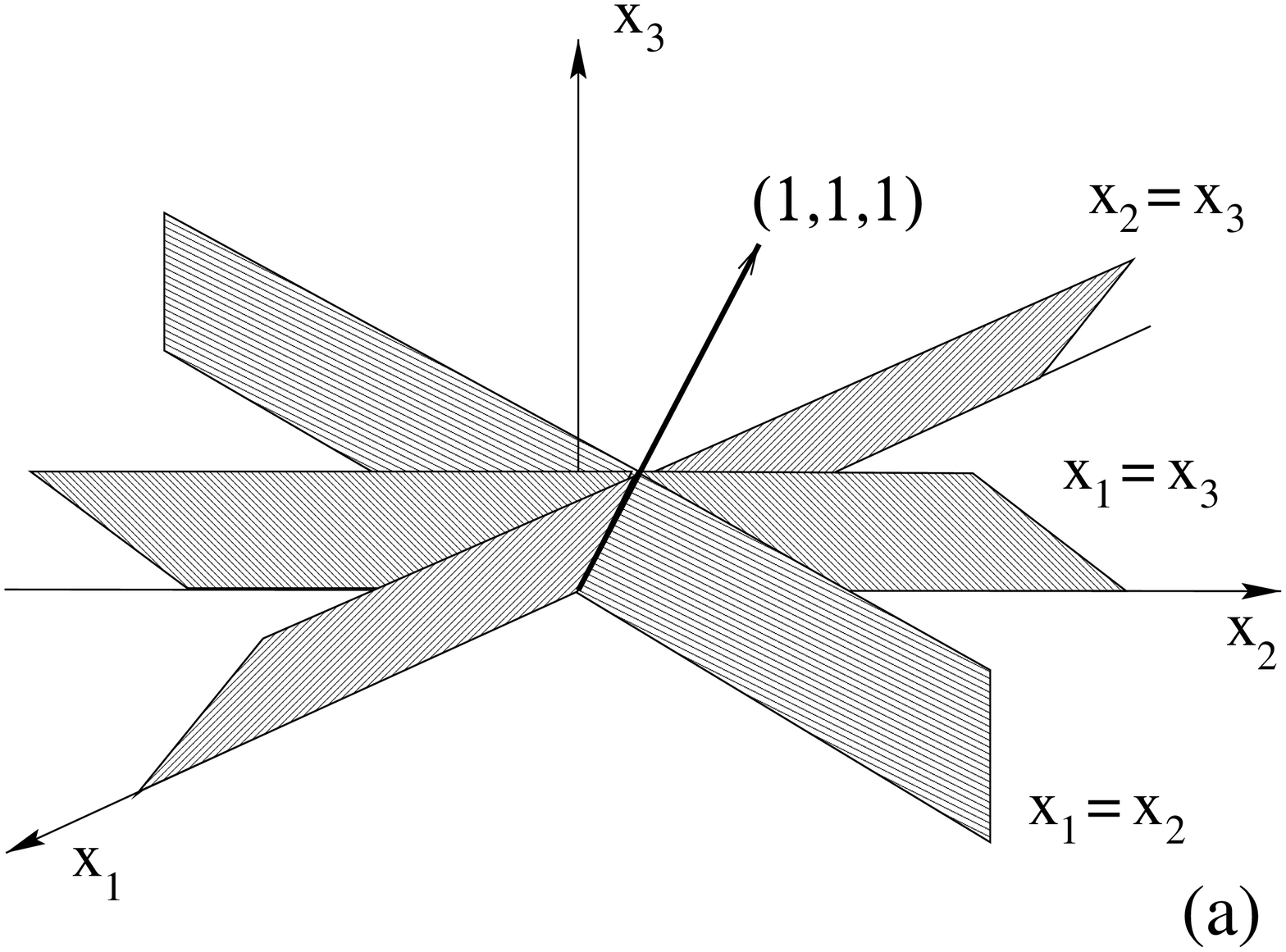}
 \includegraphics*[width=0.3\textwidth]{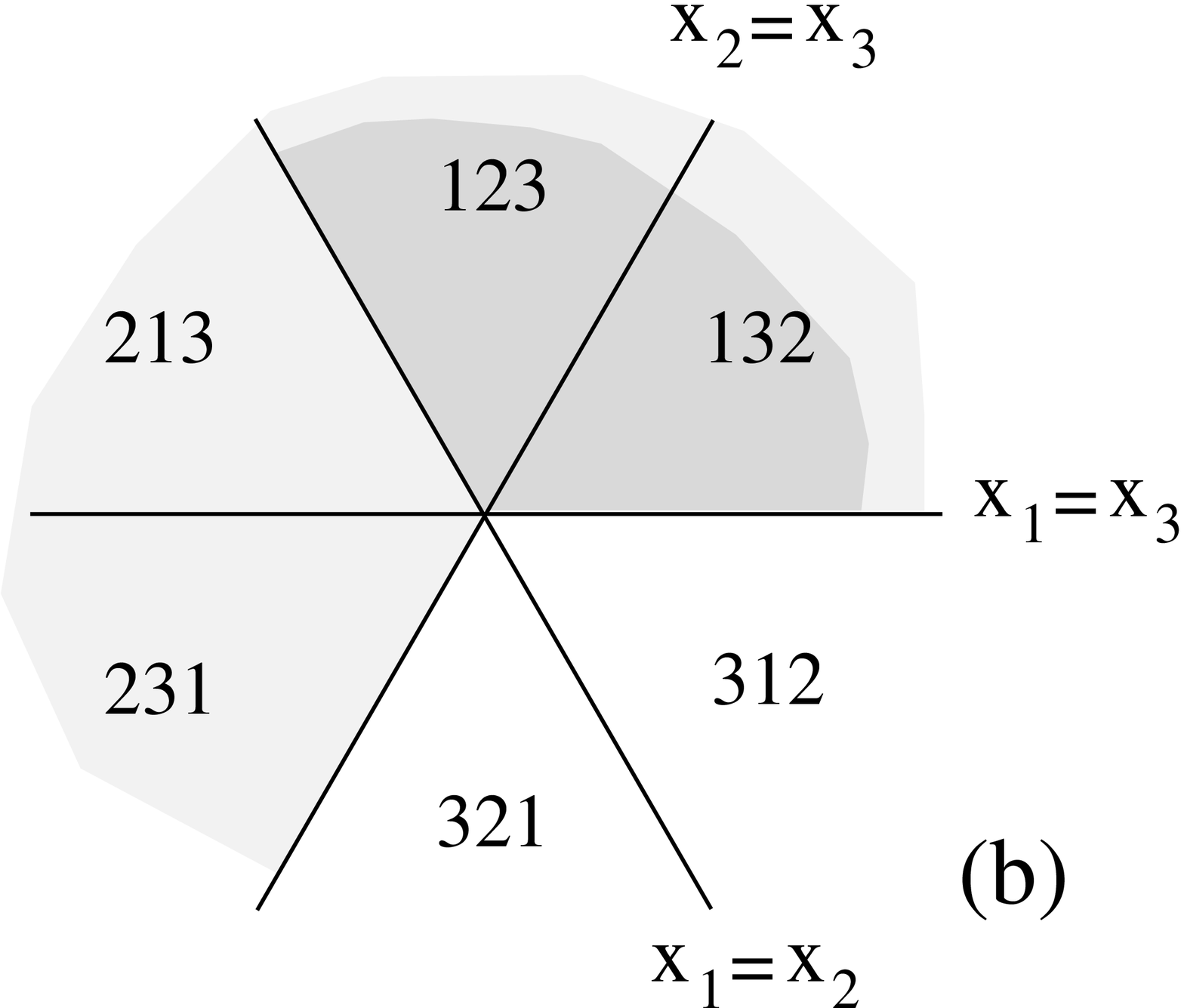}
\caption{The order domains for 3 particles in (a) the full
  3-dimensional space, and (b)~projected onto the subspace perpendicular to
  the $(1,1,1)$ axis.  The notation $ijk$ is shorthand for $x_i<x_j<x_k$.
  The allowed region corresponding to survival of a leader ($x_1<x_2, x_3$)
  is indicated by the darker shading, while the lighter shaded region
  corresponds to the laggard problem ($x_3\not < x_1,x_2$).   }
\label{planes}
\end{figure}

For 2 particles, their separation undergoes simple diffusion and the process
terminates when the separation equals zero.  Thus the survival probability of
the leader decays as $t^{-1/2}$.  To fix notation and ideas for later
sections, we now study the 3-particle system.  For a leader at $x_1(t)$ and
particles at $x_2(t)$ and $x_3(t)$, we view these coordinates as equivalent
to the isotropic diffusion of a single effective particle at
$\big(x_1(t),x_2(t),x_3(t)\big)$ in three dimensions.  Whenever this
effective particle crosses the plane $A_{ij}\!:\,x_i=x_j$, the original
walkers at $x_i$ and $x_j$ in one dimension have reversed their order.  There
are three such planes $A_{12}, A_{13}, A_{23}$ that divide space into $6$
domains, corresponding to the $3!$ possible orderings of the three walkers
(Fig.~\ref{planes}(a)).  These planes all intersect along the $(1,1,1)$ axis.

We may simplify this description by projecting onto the plane $x_1+x_2+x_3=0$
that contains the origin and is perpendicular to the $(1,1,1)$ axis.  Now the
plane $A_{12}$ may be written parametrically as $(a,a,b)$ and its
intersection with the plane $x_1+x_2+x_3=0$ is the line $(a,a,-2a)$.
Likewise, the intersections of $A_{13}$ and $A_{23}$ with the plane
$x_1+x_2+x_3=0$ are $(a,-2a,a)$ and $(-2a,a,a)$, respectively
(Figs.~\ref{planes}(b)).

The survival of the leader corresponds to the effective particle remaining
within the adjacent domains 123 and 132 in Fig.~\ref{planes}(b).  The
background particles at $x_2$ and $x_3$ are allowed to cross, but the leader
at $x_1$ always remains to the left of both $x_2$ and $x_3$.  The union of
these two domains defines a wedge of opening angle $120^{\circ}$.  Since the
survival probability of a diffusing particle within a wedge of arbitrary
opening angle $\varphi$ and absorbing boundaries decays asymptotically as
$t^{-\pi/2\varphi}$ \cite{CJ59}, we deduce the known result that the leader
survival probability exponent is $\beta_3=3/4$.

\section{Electrostatic Formulation}

For more than 3 particles, the domain for the effective particle is
geometrically more complex and the corresponding solution to the diffusion
equation does not seem tractable.  We therefore recast the survival
probability of the effective diffusing particle in terms of the simpler
problem of the electrostatic potential of a point charge within the same
geometric domain \cite{Re01}.  Let $S(t)$ be the survival probability of a
diffusing particle within an infinite wedge-shaped $d$-dimensional domain
with absorbing boundary conditions.  Let $V(r)$ be the electrostatic
potential due to a point charge within the same domain, with $V=0$ on the
boundary.  Generically, these two quantities have the asymptotic behaviors
\begin{eqnarray}
\label{asymp}
&&S(t)\propto t^{-\b}\;,\qquad t\to\infty\;,\nonumber\\
&&V(r)\propto r^{-\mu}\;,\qquad r\to\infty\;.
\end{eqnarray}

More relevant for our purposes, these two quantities are simply related by
\cite{Kr96,Re99,Re01}
\[
\int_0^t S(t)\,dt\sim\int^{\sqrt{t}}V(r)\,r^{d-1}dr\;.
\]
This equivalence arises because the integral of the diffusion equation over
all time is just the Laplace equation.  Thus the time integral of the
survival probability has the same asymptotic behavior as the spatial integral
of the electrostatic potential over the portion of the domain that is
accessible by a diffusing particle up to time $t$.  Substituting the
respective asymptotic behaviors from Eqs.~(\ref{asymp}), and noting that the
allowed wedge domain for an $N$-particle system has dimension $N-1$, we
obtain the fundamental exponent relation
\begin{equation}
\label{beta_mu}
\b=\frac{\mu-N+3}{2}\;.
\end{equation}
Thus the large-distance behavior of the electrostatic potential within a
specified domain with Dirichlet boundary conditions also gives the long-time
survival probability of a diffusing particle within this domain subject to
the same absorbing boundary conditions.  From this survival probability, we
can then determine the original ordering probability.

To illustrate this approach, let us determine the various ordering
probabilities of 3 particles on the line in terms of the equivalent
electrostatic problem.  In fact, it is simpler to work backwards and find the
equivalent ordering problem that corresponds to a specific wedge domain, For
example, consider the $60^\circ$ wedge 123 in Fig.~\ref{planes}.  If the
effective particle remains within this wedge, the initial particle ordering
on the line is preserved.  This corresponds to the vicious random walk
problem in which no particle crossings are allowed.  To obtain the asymptotic
behavior of the potential of a point charge interior to this wedge, let us
assume that $V({\bf r}) \sim r^{-\mu}f(\varphi)$ as $r\to\infty$.
Substituting this ansatz into the 2-dimensional Laplace equation, we obtain
the eigenvalue equation $f''(\varphi)=-\mu^2f(\varphi)$, subject to
$f(\varphi)=0$ on the wedge boundaries.  For the $60^\circ$ wedge, the
solution with the smallest eigenvalue is $f(\varphi)=\sin(3\varphi)$.  Thus
$\mu=3$, leading to the known result $\b=(\mu-N+3)/2=3/2$, for 3 vicious
walkers.

\section{4-Particle System} 

\begin{figure}[ht] 
 \vspace*{0.cm}
 \includegraphics*[width=0.35\textwidth]{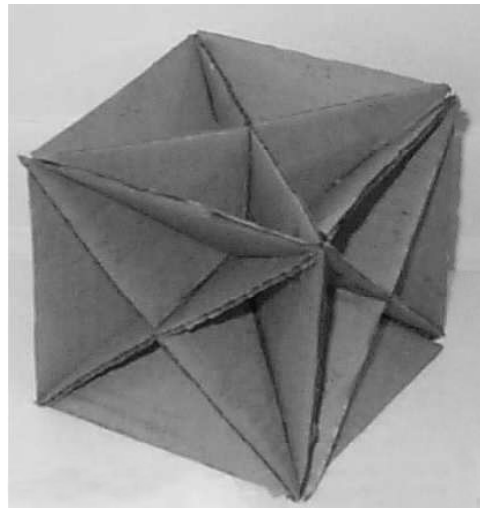}
\medskip\vspace{0.5cm}
 \includegraphics*[width=0.47\textwidth]{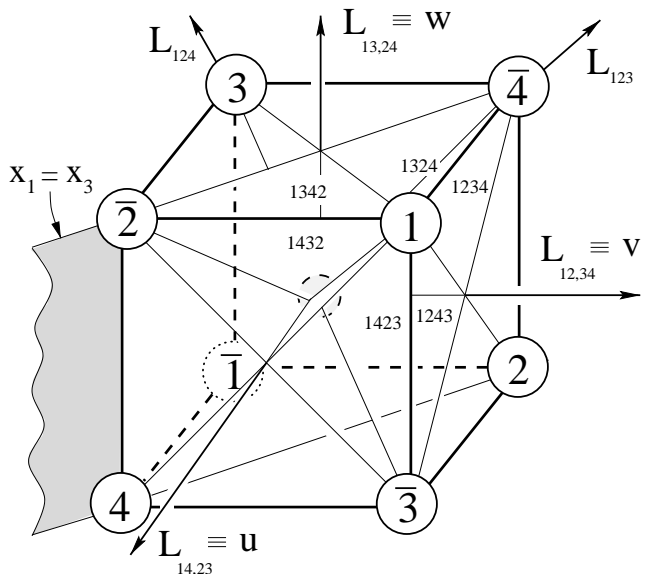}
\caption{Top: Cardboard model of the ordering domains for 4 particles on the 
  line after projection into the subspace perpendicular to $(1,1,1,1)$.  This
  structure consists of only 6 intersecting planes.  Each plane bisects the
  cube and is defined by the equality of two coordinates.  Lower: Schematic
  of the same system.  The wedge $ijkl$ denotes the region where
  $x_i<x_j<x_k<x_l$.  The union of six such ordering wedges, corresponding to
  the leader problem $x_1<x_2,x_3,x_4$, are labeled.  This domain is bounded
  by the rays $0\overline{2}$, $0\overline{3}$, $0\overline{4}$, where the
  origin is the dashed circle at the intersection of the axes
  $L_{14,23}\equiv u$, $L_{12,34}\equiv v$, and $L_{13,24}\equiv w$.  One
  constraint plane, $x_1=x_3$, is shown shaded (outside the cube).  }
\label{cube}
\end{figure}

The state of the system may be represented by an effective diffusing particle
in 4 dimensions.  By projecting onto the 3-dimensional subspace
$x_1+x_2+x_3+x_4=0$ that is orthogonal to the $(1,1,1,1)$ axis, the order
domains of the original particles can be reduced to 3 spatial dimensions.  In
this 3-dimensional subspace, the boundary $A_{12}\!:\,x_1=x_2$ becomes the
plane $(a,a,b,c)$, with $2a+b+c=0$.  Likewise, $A_{13}$ may be written
parametrically as $(a,b,a,c)$, where again $2a+b+c=0$.  Thus the locus
$L_{123}\equiv A_{12}\cap A_{13}\!:\,x_1=x_2=x_3$ is the line $(a,a,a,-3a)$.
This body diagonal joins the nodes 4 and $\overline{4}$ in Fig.~\ref{cube}.
Along this axis, the original particle coordinates on the line obey the
constraint $x_1=x_2=x_3$, with $x_4<x_3$ on the half-axis closer to node 4
and $x_4>x_3$ on the half-axis closer to node $\overline{4}$.  A similar
description applies to the axes $L_{124}=(a,a,-3a,a)$ between 3 and
$\overline{3}$, $L_{134}=(a,-3a,a,a)$ between 2 and $\overline{2}$, and
$L_{234}=(-3a,a,a,a)$ between 1 and $\overline{1}$.

The locus where $x_1=x_2$ and $x_3=x_4$ simultaneously, is the line
$L_{12,34}\equiv A_{12}\cap A_{34}=(a,a,-a,-a)$.  Likewise,
$L_{13,24}=(a,-a,a,-a)$, and $L_{14,23}=(a,-a,-a,a)$.  Viewed in the
orthogonal 3-subspace, the $6$ planes $A_{ij}$ intersect in $7$ lines ($4$
$L_{ijk}$ and $3$ $L_{ij,kl}$), and divide the subspace into $24$
semi-infinite wedges, as shown in Fig.~{cube}(b).  Each of these wedges
corresponds to one of the $4!$ orderings of the walkers in one dimension.

We first illustrate the electrostatic formulation of this system for the
vicious random walk problem.  Since the initial particle order is preserved,
the effective diffusing particle remains within a single wedge $ijkl$ in
Fig.~\ref{cube}.  As outlined in the previous section, the survival
probability of this effective particle corresponds to the electrostatic
potential of a point charge within this one wedge, with the boundary surfaces
held at zero potential.  To solve this electrostatic problem, it is
convenient to place the point charge at the symmetric location
$(u,v,w)=(0,1,1/2)$ within the wedge 1234, where the $(u,v,w)$ axes are
defined in Fig.~\ref{cube}.  From the image method, the potential due to
a point charge within one wedge is equivalent to the potential of an array of
24 symmetrically-placed point charges consisting of the initial charge and 23
image charges, with positive images at $-(0,1,1/2)$, $\pm(0,1,-1/2)$,
$\pm(\pm 1/2,0,1)$, $\pm(1,\pm 1/2,0)$, and negative images at
$\pm(\pm 1/2,1,0)$, $\pm(0,\pm 1/2,1)$, and $\pm(0,1,\pm 1/2)$.  Using
Mathematica the asymptotic behavior of the potential in wedge 1234 (where the
original charge is placed) due to this charge array is
\[
V(r) = a_1r^{-7} + a_2r^{-11} + a_3r^{-13} + a_4r^{-15} +\cdots\;.
\]
(The coefficients $a_i$ depend on the location of the charge and on the
orientation of {\bf r}.)  Using the exponent relation~(\ref{beta_mu}), the
asymptotic survival probability of $4$ vicious walkers is given by
\[
S(t) = b_1t^{-3} + b_2t^{-5} + b_3t^{-6} + b_4t^{-7} +\cdots\;.
\]
The leading behavior confirms the known result \cite{Fi84,HF84}, and we
obtain the form of the corrections as well.  As an amusing aside, notice that
an $r^{-7}$ dependence for the potential is normally achieved by a 64-pole
charge configuration.  The high symmetry of the 24 charges in the ordering
problem leads to a multipole field normally associated with at least 64 point
charges.

We next turn to the leader problem.  This system corresponds to the
electrostatic problem within the combined domain of the 6 wedges marked 1234,
1243, 1423, 1432, 1342, and 1324, in Fig.~\ref{cube}.  The resulting
domain is a tetrahedral corner, with its apex at the center of the cube, that
is flanked by the rays $0\overline{2}$, $0\overline{3}$, and $0\overline{4}$.
Despite the simplicity of this domain, we are unable to solve this
electrostatic problem analytically and we have instead studied the problem
numerically.

In the allowed region of the cube in Fig.~\ref{cube}, we discretize
space and solve the electrostatic potential of a point charge by using
successive over-relaxation, with the domain boundaries at zero potential.
For simplicity, the charge is chosen to be at the symmetric point $(1,1,1)$.
For the outer faces of the cube we use two different boundary conditions:
(a)~absorbing ($V=0$), and (b)~reflecting, ($dV/dn=0$, where $n$ is the
direction normal to the surface).  The true potential -- that of the infinite
wedge -- lies between these two extremes.  We also exploit the symmetry about
the $(1,1,1)$ diagonal and use only the domain marked 1234 in
Fig.~\ref{cube}, with absorbing boundary conditions on the plane
$34\overline{3}\overline{4}$, and reflecting boundary conditions on the
planes $12\overline{1}\overline{2}$ and $14\overline{1}\overline{4}$.  As
already discussed, the outer cube face, $1\overline{3}2\overline{4}$, is
taken to be reflecting or absorbing.  This space savings allows us to carry
out computations for a cube of $500$ lattice spacings on a side.

\begin{figure}[ht] 
 \vspace*{0.cm}
 \includegraphics*[width=0.45\textwidth]{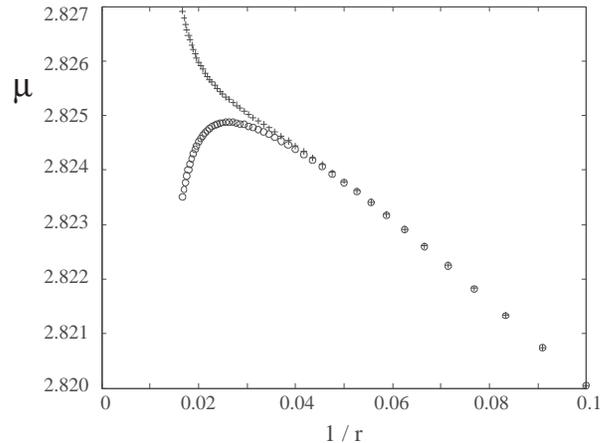}
\caption{Local exponent $\mu(r)$, as a function of $1/r$ for the
  tetrahedral wedge with absorbing $(+)$ and reflecting $(\circ)$ finite-size
  boundary conditions at the outer face of the cube. }
\label{local-exp}
\end{figure} 

While the power-law decay of the potential sets in quickly, the finite-size
effect is pronounced and it is perceptible already at 25 lattice spacings
away from the charge.  This is the primary limitation on the accuracy of our
exponent estimate.  Nevertheless, the local exponent $\mu(r)=-d\ln V(r)/d\ln
r$ varies only at the fourth digit (Fig.~\ref{local-exp}).  The approach of
the local exponent to the asymptotic limit also suggests that $V(r)$ has the
form $V(r)\sim r^{-\mu}+Ar^{-4}$.  Assuming that this is the case,
extrapolation of the data in Fig.~\ref{local-exp} gives
$\mu=2.82684\pm0.00016$, where the error bar is the difference in the
extrapolated value of $\mu(r)$ from the two different boundary conditions.
{}From the exponent correspondence given in Eq.~(\ref{beta_mu}), we thereby
obtain, for the lead probability,
\begin{equation}
\label{b4}
{\cal L}_4(t)\sim t^{-\beta_4}+At^{-3/2}\;,\qquad \beta_4=0.91342(8)\;,
\end{equation}

\begin{figure}[ht] 
 \vspace*{0.cm}
 \includegraphics*[width=0.45\textwidth]{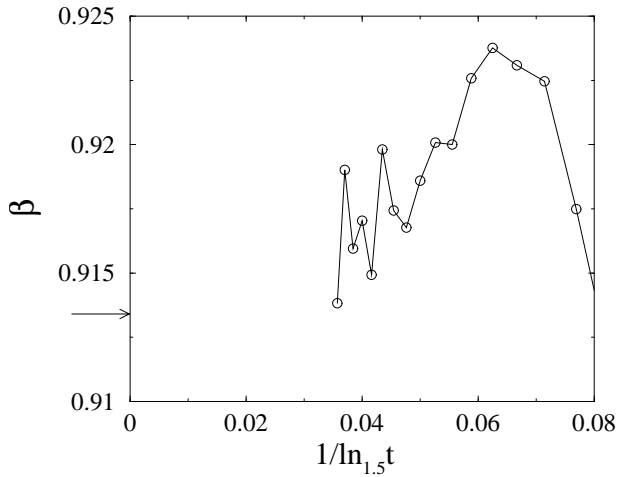}
\caption{Direct simulation results for the local exponent in the survival
  probability for $10^9$ configurations.  The arrow indicates our estimate
  for $\beta_4$ from Eq.~(\ref{b4}).}
\label{beta-direct}
\end{figure}

It is hard to match this numerical accuracy with that from direct simulations
of the survival of the leader.  We simulated $10^9$ realizations of the
system; this gives extremely linear data for the time dependence of the
leader survival probability on a double logarithmic scale.  To estimate the
exponent $\beta_4$, we computed the local slopes of the survival probability
versus time in contiguous time ranges between $t$ and $1.5t$ when plotted on
a double logarithmic scale.  These local exponents are plotted against
$1/\ln_{1.5}t$ (Fig.~\ref{beta-direct}).  The results are compatible but much
less accurate than Eq.~(\ref{b4}).

\section{The Laggard Problem}

In the laggard problem, we study the probability that the initially rightmost
particle at $x_{N}$ has never been the leader during the time interval
$(0,t)$.  The laggard problem can also be recast into the diffusion of a
single effective particle within an $N$--dimensional wedge-shaped region,
with absorbing domain boundaries.  This mapping leads to the basic conclusion
that every particle that is initially not in the lead exhibits the same
asymptotic behavior as the last particle.  Indeed, for any particle $i$
initially at $x_i$, the regions $x_i\not <x_1,\ldots, x_{i-1},
x_{i+1},\ldots,x_{N}$ are isomorphic.  The initial condition merely fixes the
location of the effective particle in this allowed region.  Another
fundamental observation is that the allowed regions of the effective particle
for the leader and the laggard problems are {\em complementary} for all $N$.

For two particles, the probability that the laggard does not become the
leader obviously decays as $t^{-1/2}$, {\it i.e.}, $\gamma_2=1/2$.  The case
$N=3$ is more interesting but also solvable.  The condition that a particle
initially at $x_3$ never attaining the lead ($x_3\not<$ $x_1,x_2$) is
equivalent to the effective particle remaining with the lighter shaded region
in Fig.~\ref{planes}(b).  Since the opening angle of the resultant wedge is
$\varphi=4\pi/3$, the corresponding survival probability decays as
$t^{-3/8}$, implying that $\gamma_3=3/8$.

We have performed direct numerical simulations of the process to
estimate the exponent $\gamma_N$ for $N=3, 4,5$, and 6.  Each simulation is
based on $10^6$ realizations in which $N-1$ particles are initially at the
origin, while the laggard is at $x=1$.  Each simulation is run until the
laggard achieves the lead or $10^5$ time steps, whichever comes first.  From
the survival probability, we estimate $\gamma_N\approx 0.35, 0.30, 0.26$, and
$0.23$ for $N=3$ -- 6, respectively.  Since we know that $\gamma_3=0.375$,
the discrepancy of 0.025 between the simulation result and the theory is
indicative of the magnitude of systematic errors in this straightforward
numerical approach.

While it appears difficult to determine the exponent $\gamma_N$ analytically
for general $N>3$, the situation simplifies in the large $N$ limit because
the position of the leader becomes progressively more deterministic.  Indeed,
the probability density $P_N(y,t)$ that the leader is located at distance $y$
from the origin (assuming that all particles are initially at the origin) is
\cite{Re99}
\begin{eqnarray*}
\label{Ln}
P_N(y,t)={N\, e^{-y^2/4Dt}\over \sqrt{4\pi D t}}
\left[1-{1\over 2}\,{\rm erfc}\left(-{y\over \sqrt{4Dt}}\right)\right]^{N-1},  
\end{eqnarray*}
where ${\rm erfc}(x)=1-{\rm erf}(x)$ is the complementary error function.
Performing a large $N$ analysis, we find that the probability density
$P_N(y,t)$ approaches a Gaussian
\begin{equation}
\label{Lnasymp}
P_N(y,t)\to {1\over \sqrt{2\pi \sigma^2}}\,
\exp\left\{-{(y+y_*)^2\over 2\sigma^2}\right\},
\end{equation}
with the mean and the variance of this distribution given by
\begin{equation}
\label{ysigma}
y_*=z\sqrt{4Dt}\,, \qquad 
\sigma^2={Dt\over z^2}\,,
\end{equation}
where $z$ is determined from the transcendental relation
\begin{equation}
\label{z}
z\,e^{z^2}={N\over \sqrt{4\pi}}. 
\end{equation}
Consequently, the parameter $z$ diverges as $z\approx \sqrt{\ln N}$ when
$N\to\infty$.  The ratio of the dispersion to the mean displacement thus
decreases as $\sigma/y_*\approx (2\ln N)^{-1}$, so that the position of the
leader indeed becomes more deterministic as $N\to\infty$.

Therefore in the large $N$ limit we can assume that the leader is moving
deterministically and its position is given by $-y_*(t)$.  Then the
probability that the laggard never achieves the lead is equivalent to the
probability that a diffusing particle initially at the origin does not
overtake a receding particle whose position is varying as $-y_*(t)$.  This
corresponds to the solution to the diffusion equation in the expanding region
$x\in(-y_*(t),\infty)$ with an absorbing boundary condition at the receding
boundary $x=-y_*$.  When $y_*\propto t^{1/2}$, this diffusion equation can be
solved exactly by reducing it to a parabolic cylinder equation in a fixed
region.  However, in the limit $t\to\infty$, we can obtain asymptotically
correct results much more simply.  Because the absorbing boundary recedes
from the laggard particle relatively quickly, we solve the problem by
assuming that the density $P(x,t)$ of the laggard particle approaches a
Gaussian with yet unknown amplitude ${\cal R}_N(t)$ \cite{Kr96,Re99}:
\begin{equation}
\label{Pn}
P(x,t)={{\cal R}_N(t)\over \sqrt{4\pi D t}}\,
\exp\left\{-{x^2\over 4Dt}\right\}.  
\end{equation}
Although this distribution does not satisfy the absorbing boundary condition,
the inconsistency is negligible since the exponential term in Eq.~(\ref{Pn})
is of order $N^{-1}$ at the boundary $y=y_*$.  

The probability ${\cal R}_N(t)$ is now found self-consistently by equating the
``mass'' loss to the flux:
\begin{equation}
\label{Pnrel}
{d {\cal R}_N\over dt}=D{\partial P\over \partial x}\Big|_{x=-y_*}
\end{equation}
Using Eq.~(\ref{Pn}) to compute the flux we convert Eq.~(\ref{Pnrel}) into
\begin{equation}
\label{Pndif}
{d {\cal R}_N\over dt}=-{z^2\over N}\,{{\cal R}_N\over t},
\end{equation}
{}from which the exponent $\gamma_N$ is $z^2/N$.  This is of course valid
only in the large $N$ limit.  Taking this limit in Eq.~(\ref{z}) we obtain
\begin{equation}
\label{psin} 
\gamma_N={\ln N\over N}-{\ln\ln N\over 2N}+\ldots .
\end{equation}
Thus as $N$ gets large, the probability that the laggard never attains the
lead decays extremely slowly with time.  This fits with the naive intuition
that if the number of particles is large a laggard initially is very likely
to remain a laggard.  Each additional particle makes it even less likely that
the laggard could achieve the lead.  Amusingly, this asymptotic exponent
predictions is numerically close to the previously-quoted results from direct
numerical simulations of the laggard problem.

\section{Concluding Remarks}

We investigated two dual random walk ordering problems in one dimension: (i)
what is the probability that a particle, that is initially in the lead,
remains in the lead and (ii) what is the probability that a particle, that is
initially not in the lead, never achieves the lead?  These problems are most
interesting in one spatial dimension because of the effective correlations
between the interacting particles.  These correlations are absent in two
dimensions and greater, so that an $N$-particle system reduces to a
2-particle system \cite{Kr96,Re99,CK02}.

We determined the respective exponents $\beta_N$ and $\gamma_N$ associated
with the lead and laggard probabilities for general $N$.  Both exponents can
be determined by elementary geometric methods for $N=2$ and 3 and by
asymptotic arguments for $N\to\infty$.  Our new results are the following:
(i) a precise estimate for $\beta_4$ and (ii) the large-$N$ behavior of
$\gamma_N$.

A simple generalization is to allow each particle $i$ to have a distinct
diffusion coefficient $D_i$.  The exponents $\beta_N$ and $\gamma_N$ will now
depend on the diffusion coefficients, except for $N=2$, where $\beta_2$ and
$\gamma_2$ always equal 1/2.  The case $N=3$ is still solvable by introducing
rescaled coordinates $y_i=x_i/\sqrt{D_i}$ to render the diffusion of the
effective particle isotropic, after which the mapping to the wedge can be
performed straightforwardly.  We thus find
\begin{eqnarray*}
\beta_3&=&\left\{2-{2\over\pi}\,
\cos^{-1} {D_1\over\sqrt{(D_1+D_2)(D_1+D_3)}}\right\}^{-1}\;,\\
\gamma_3&=&\left\{2+{2\over\pi}\,
\cos^{-1} {D_3\over\sqrt{(D_1+D_3)(D_2+D_3)}}\right\}^{-1}\;.
\end{eqnarray*}
An amusing special case is the case of a stationary laggard, for which
$\gamma_3=1/3$.  

Finally, it is also worth mentioning a promising development to solve the
diffusion equation in the domains defined by the ordering of one-dimensional
random walks.  This is the recent discovery of deep connections between
vicious walkers and random matrix theory \cite{G99,Baik00,KT02}.  These allow
one to not only re-derive the exponent $\alpha_N$ of the original vicious
random walk problem, but also lead to many new results.  It would be extremely
useful if these techniques could be extended to the leader and laggard
problems.

\acknowledgments

We are grateful to NSF grants PHY-0140094 (DbA) and DMR9978902 (PLK and SR)
for partial financial support of this research.

%%%%%%%%%%%%%%%%%%%%%%%%%%%%%%%%%% REFERENCES %%%%%%%%%%%%%%%%%%%%%%%%%


\begin{thebibliography}{99}
\bibitem{remark} We assume that all random walkers have the same
  diffusion coefficient $D$ if not stated otherwise.  For distinct
  diffusivities $D_i$ for each particle, one may rescale each coordinate by
  $x_i\to y_i=x_i/\sqrt{D_i}$ so that the motion of the effective particle is
  again isotropic.
\end{thebibliography}
\end{document}